\documentclass[11pt]{article}
\usepackage[utf8]{inputenc}
\usepackage[T1]{fontenc}
\usepackage[margin=1in]{geometry}
\usepackage{amsmath,amssymb}
\usepackage{authblk}
\usepackage{graphicx}
\usepackage{enumitem}
\usepackage{xcolor}
\usepackage{hyperref}
\hypersetup{colorlinks=true,linkcolor=blue!50!black,citecolor=blue!50!black,urlcolor=blue!50!black}
\usepackage{listings}
\usepackage{tikz}
\usetikzlibrary{arrows.meta,positioning,fit,backgrounds}
\usepackage{pgfplots}
\pgfplotsset{compat=1.18}
\usepackage{booktabs}
\usepackage{threeparttable}
\usepackage{multirow}
\usepackage{array}
\usepackage{titlesec}
\titlespacing*{\section}{0pt}{1.4em}{0.7em}
\titlespacing*{\subsection}{0pt}{1.1em}{0.5em}
\title{\textbf{TPM-Attest: Hardware-Rooted Integrity Attestation as a Kernel-Level Anti-Cheat Alternative for Linux}}

\author[1]{Anudeep Gedela}
\author[1,*]{Dr.\ A.\ Yaswanth}
\affil[1]{GITAM School of Technology, GITAM (Deemed to be University), Visakhapatnam, India}
\affil[*]{Corresponding author -- Assistant Professor, Department of Computer Science \& Engineering, \texttt{yamanapu@gitam.edu}}
\date{}

\begin{document}
\maketitle

\begin{abstract}
Multiplayer PC gaming on Linux faces a structural problem: the anti-cheat systems that publishers require operate as proprietary Ring~0 kernel modules that are architecturally incompatible with Linux's security model, GPL licensing, and stable ABI guarantees. We argue the right response is not to port these invasive modules to Linux, but to replace them entirely. This paper presents \textbf{TPM-Attest}, a hardware-rooted remote attestation framework that uses the Trusted Platform Module (TPM)~2.0 and the Linux Integrity Measurement Architecture (IMA) to prove, cryptographically, that a client booted cleanly and ran only authorised software -- without any kernel driver, without proprietary code, and without scanning player memory. The system intercepts Epic Online Services (EOS) SDK calls via a userspace \texttt{LD\_PRELOAD} hook, gates session access on a live TPM quote bound to a server-issued nonce, and constructs an index-prefixed Merkle tree over the IMA log that is immune to duplicate-leaf collision attacks. Across 500 constructed tamper sessions we achieve a 100\% detection rate; incremental leaf caching reduces repeat-attestation latency to under 3~seconds on real TPM~2.0 hardware. A controlled red-team evaluation against a live demo game confirms all four file-backed attack vectors are blocked while precisely characterising the two confirmed bypass conditions. The full implementation is released as open-source software.
\end{abstract}

\section{Introduction}
\label{sec:intro-ref}

Valve's Proton compatibility layer has made thousands of Windows titles playable on Linux~\cite{valve2018proton}, yet competitive multiplayer games remain off-limits for most Linux players. The barrier is not the game logic -- it is the anti-cheat. Systems like Easy Anti-Cheat and BattlEye rely on Ring~0 kernel drivers that assume a Windows environment, conflict with Linux's stable ABI guarantees, and cannot be shipped under GPL. The result is straightforward: Linux players are locked out.

Worse, this lockout carries a hidden cost for everyone. A recent audit of four widely deployed kernel-mode anti-cheat systems found that two behaved indistinguishably from rootkits -- broad OS visibility, covert channels, and access far exceeding any legitimate integrity-checking purpose~\cite{dorner2024rootkit}. Players on Windows accept these risks because they have no choice; Linux's kernel rejects such code, which is why Linux support is absent.

We take a different approach. Instead of scanning memory at runtime, TPM-Attest asks one question at session startup: \textit{did this machine boot into a known-good state and execute only authorised software?} The TPM~2.0 microcontroller, present on virtually every modern motherboard, can answer this with a hardware-signed cryptographic quote~\cite{tcg2018}. Paired with the Linux Integrity Measurement Architecture -- a kernel subsystem maintaining a tamper-evident log of every executed binary and loaded library since boot~\cite{sailer2004ima} -- the result is an attestation report the server can verify in under 15 seconds. A clean chain permits the session; any deviation denies access before the game reaches the network~\cite{rats2023}.

This paper makes five core contributions:
\begin{enumerate}[nosep]
  \item An open-source, end-to-end TPM-Attest framework combining TPM~2.0 quotes and IMA measurement logs for Linux gaming integrity verification.
  \item An index-prefixed Merkle tree construction immune to duplicate-leaf collision attacks (CVE-2012-2459 class).
  \item A dynamic \texttt{LD\_PRELOAD} interception hook gating EOS SDK sessions on real-time attestation results.
  \item A comprehensive experimental evaluation including latency profiling, tamper-detection rates, and a comparative analysis against Keylime and kernel-mode anti-cheat baselines.
  \item \textbf{VOID SECTOR} -- a complete playable 2-D space-shooter demo game whose session launch is gated end-to-end by the TPM-Attest pipeline, providing a live, reproducible demonstration of the full attestation flow (game client $\to$ hook $\to$ daemon $\to$ TPM $\to$ server $\to$ session granted/denied).
\end{enumerate}

\section{Background and Foundations}

\subsection{TPM 2.0 and Platform Configuration Registers (PCRs)}
The TPM~2.0 microcontroller operates as a secure cryptoprocessor physically isolated from the host CPU~\cite{tcg2018}. Platform Configuration Registers (PCRs) are volatile memory slots storing hashes representing the platform's boot and software state, structured in accordance with NIST BIOS integrity measurement guidelines~\cite{nist800155}. PCRs are updated via a one-way \textbf{extend} operation:

\begin{equation}
  \text{PCR}_{\text{new}} = \text{SHA-256}\!\left(\text{PCR}_{\text{old}} \,\|\, m\right)
  \label{eq:pcr-extend}
\end{equation}

\noindent where $m$ is the new measurement value. This one-way chaining prevents register rollback or state forging. To prove register contents to an external verifier, the TPM provides the quote operation (\texttt{tpm2\_quote}). The microcontroller signs a composite digest over designated PCRs using its internal, non-migratable Attestation Key (AK), binding a server-supplied challenge nonce $\eta$ directly into the signed structure to defeat replay attacks~\cite{tpm2tools2024}. Freshness is cryptographically guaranteed.

\subsection{Integrity Measurement Architecture (IMA)}
Alongside the hardware cryptoprocessor, the Linux Integrity Measurement Architecture (IMA) hooks the virtual filesystem to record software execution at the kernel level~\cite{lko2021}. Whenever an executable binary or shared library is mapped into memory via \texttt{execve} or \texttt{mmap}, IMA intercepts the system call, computes the cryptographic digest of the file contents, logs the record into an append-only ASCII event table, and immediately extends the measurement into PCR~10. The chain begins at boot. To link runtime executions back to the physical firmware state, IMA anchors the log with a dedicated \texttt{boot\_aggregate} measurement over PCRs 0--7:

\begin{equation}
  \texttt{boot\_aggregate} = \text{SHA-256}\!\left(\bigoplus_{i=0}^{7} \text{PCR}_i\right)
  \label{eq:boot-agg}
\end{equation}

\subsection{Shared Library Interception (\texttt{LD\_PRELOAD})}
Shared library preloading is a standard Unix mechanism~\cite{kerrisk2010}. By configuring the \texttt{LD\_PRELOAD} environment variable, the dynamic linker prioritises our shim library ahead of standard system dependencies. We export exact matching function signatures for critical Epic Online Services (EOS) SDK routines, intercept outbound network attempts before packet transmission, query local attestation state over a Unix socket, and drop unauthorised calls in-process. No game source code modification is required.

\subsection{Index-Prefixed Merkle Tree}
Naive binary Merkle trees have a well-documented vulnerability: duplicate-leaf collisions~\cite{bitcoin2012cve}. If an unauthenticated log re-orders entries or appends duplicate measurements, a standard tree can generate identical intermediate hashes, allowing an adversary to bypass log completeness checks. TPM-Attest prevents this vulnerability entirely. We prepend the 64-bit absolute log index $i$ directly to the leaf hash preimage:

\begin{equation}
  \ell_i = \text{SHA-256}\!\left(i \,\|\, \text{pcr}_i \,\|\, \text{tmpl\_hash}_i \,\|\, \text{file\_hash}_i \,\|\, \text{filename}_i\right)
  \label{eq:leaf}
\end{equation}

Internal nodes are computed as:
\begin{equation}
  n_{\text{parent}} = \text{SHA-256}\!\left(\ell_{\text{left}} \,\|\, \ell_{\text{right}}\right)
  \label{eq:internal}
\end{equation}

Position binding is absolute. Because each leaf hash $\ell_i$ explicitly embeds its sequence index $i$, reordering, duplicating, or truncating log entries inevitably yields a divergent root hash that the verifier rejects on arrival.

\section{Related Work}
\label{sec:related-work}

The idea of using hardware measurements to verify software integrity on a running host dates to Sailer et al., who introduced IMA and showed that a remote verifier can reliably detect compromised binaries without modifying the client kernel~\cite{sailer2004ima}. \textbf{Keylime}, developed at MIT Lincoln Laboratory and now a CNCF-graduated project, industrialised this into a scalable agent/registrar/verifier architecture for cloud, edge, and IoT deployments~\cite{keylime}. TPM-Attest draws on the same measurement primitives but targets a different problem: gating a single consumer desktop's game session at the moment an EOS SDK call fires, rather than monitoring a server fleet continuously.

The strongest motivation for this work is Dorner and Klausner's 2024 independent forensic audit, which revealed that commercial kernel-mode drivers frequently abuse Ring~0 privileges to maintain unmonitored telemetry streams, establish covert channels, and perform sweeping memory inspection far beyond their declared anti-cheat function~\cite{dorner2024rootkit}. TPM-Attest is a direct architectural response: the trust anchor shifts into dedicated hardware and the third-party kernel driver is eliminated entirely. Alangari and Alharbi's 2025 survey of anti-cheat defences confirms that no existing Linux-native solution combines hardware-rooted trust with EOS SDK integration and GPL compatibility~\cite{alangari2025review}. Coker et al.\ formalised remote attestation properties~\cite{coker2011principles}; our nonce challenge-response design satisfies their freshness and binding criteria directly within the IETF RATS architecture~\cite{rats2023}, conforming to Entity Attestation Token (EAT) claims schemas~\cite{ietf9711}. Bratus et al.\ and Petroni et al.\ argue that passive load-time measurement and active runtime scanning are complementary~\cite{bratus2008,petroni2004} -- a perspective we adopt in Section~8 by separating platform trust from memory policing.
 
An alternative hardware trust mechanism lies in Trusted Execution Environments (TEEs) such as Intel SGX~\cite{anati2013sgx,costan2016sgx}, where isolated enclaves measure code pages and issue CPU-signed attestation tokens. However, enclaves are ill-suited to consumer gaming: they require invasive engine partitioning, incur steep cross-boundary penalties, and leave the host kernel, display server, and graphics drivers completely unmeasured~\cite{costan2016sgx}. Whole-system TPM+IMA attestation establishes platform trust across diverse Linux distributions without requiring game developers to re-architect their software.

\section{Threat Model and Security Goals}
\label{sec:threat-model}

\subsection{Adversary Capabilities}
We model a motivated adversary who has obtained root access to the target machine and seeks to gain an unfair advantage in a multiplayer game without detection. The adversary may replace or patch game binaries on disk, inject cheat libraries at process-launch time, load unsigned kernel modules, capture and replay a previously accepted attestation report, or run the game inside a virtual machine equipped with a software-emulated TPM.

\subsection{Prevented Attacks}
\begin{enumerate}[nosep]
  \item \textbf{Binary and library tampering:} If an adversary replaces or patches an executable, IMA logs the modified hash during \texttt{execve} and extends PCR~10. The resulting Merkle root diverges, triggering immediate server rejection.
  \item \textbf{Kernel modification:} Tampering with the kernel binary or boot modules alters PCR~9. The stored baseline breaks.
  \item \textbf{Report replay:} An attacker intercepting a valid quote cannot reuse it. Every attestation payload is bound to a fresh, server-generated cryptographic nonce with a 30-second expiry.
  \item \textbf{TPM emulation (vTPM/swtpm):} Virtualised or software-emulated TPMs cannot forge manufacturer-signed Endorsement Key certificates. In \texttt{STRICT\_EK\_VERIFICATION} mode, the server drops uncertified quotes on arrival.
  \item \textbf{Unauthorised MOK kernel:} Custom kernels signed with third-party Machine Owner Keys alter the Secure Boot measurement in PCR~7. The session is denied.
\end{enumerate}

\subsection{Out-of-Scope Attacks}
We explicitly define our attack boundaries. Physical bus-level attacks -- such as SPI/LPC sniffing or malicious PCIe DMA devices -- bypass operating system controls and fall outside scope. Similarly, runtime code injection via \texttt{ptrace} or anonymous mappings (\texttt{mmap}) into already-measured processes cannot be detected by load-time hooks: IMA measures file objects at invocation, never dynamic memory modified post-launch. Section~\ref{sec:redteam} confirms these boundary conditions empirically; Section~\ref{sec:limitations} outlines mitigations.

\section{System Design and Architecture}
\label{sec:design}

\begin{figure}[htbp]
\centering
\begin{tikzpicture}[
    box/.style={draw,rectangle,rounded corners,align=center,minimum height=0.70cm,inner sep=3pt,font=\small},
    >=Stealth
]
\node[box,minimum width=2.8cm] (client) {Game Client};
\node[box,below=0.38cm of client,minimum width=2.8cm] (hook) {\texttt{eac\_hook.so}\\(LD\_PRELOAD Hook)};
\node[box,right=1.8cm of hook,minimum width=2.8cm] (shim) {\texttt{shim.py}\\(Local Daemon)};
\node[box,below=0.38cm of shim,minimum width=2.8cm] (report) {\texttt{report.py}\\(Report Generator)};
\node[box,below left=0.50cm and -0.15cm of report,minimum width=2.1cm] (tpm) {TPM 2.0\\Hardware};
\node[box,below right=0.50cm and -0.15cm of report,minimum width=2.1cm] (ima) {IMA\\Security};
\draw[->] (client) -- node[right,font=\tiny]{calls EOS SDK APIs} (hook);
\draw[<->] (hook) -- node[above,font=\tiny]{Unix socket} (shim);
\draw[->] (shim) -- node[right,font=\tiny]{queries agent} (report);
\draw[->] (report) -- node[midway,left,font=\tiny]{reads PCR} (tpm);
\draw[->] (report) -- node[midway,right,font=\tiny]{reads IMA log} (ima);
\begin{scope}[on background layer]
\node[draw,dashed,inner sep=0.22cm,fit=(client)(hook)(shim)(report)(tpm)(ima),
      label={[font=\small,yshift=-2pt]above:\textbf{Attested Machine}}] (machine) {};
\end{scope}
\node[box,below=0.50cm of machine.south,minimum width=4.4cm] (server) {Verification Server\\\texttt{server/main.py}};
\draw[->] (machine.south) -- node[midway,right,font=\tiny]{HTTP POST \texttt{/attest}} (server);
\end{tikzpicture}
\caption{TPM-Attest system architecture: client-side attestation agent, local interception hook/daemon, and remote verification server.}
\label{fig:architecture}
\end{figure}
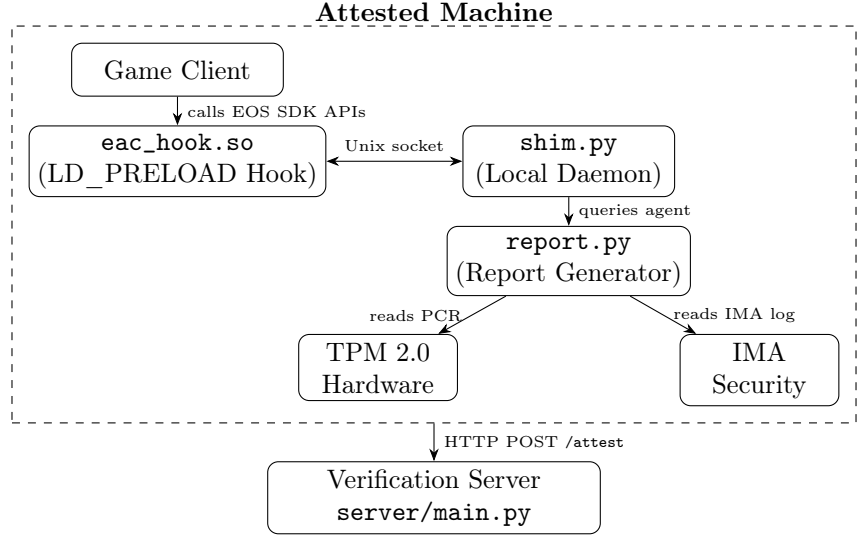

TPM-Attest is built around three loosely coupled components that together form the complete attestation pipeline (Figure~\ref{fig:architecture}). Each component communicates only through well-defined interfaces -- Unix socket or HTTP -- so any layer can be replaced or audited independently without touching the others.

\subsection{Phase 1: Client-Side Attestation Agent}
\begin{itemize}[nosep]
  \item \textbf{PCR Reader (\texttt{agent/pcr\_reader.py}):} Spawns \texttt{tpm2\_pcrread} to retrieve SHA-256 PCR bank for indices $\{0,1,4,7,9,10\}$.
  \item \textbf{IMA Reader (\texttt{agent/ima\_reader.py}):} Parses the kernel measurement log and computes the Merkle tree per Equations~\ref{eq:leaf}--\ref{eq:internal}.
  \item \textbf{Report Generator (\texttt{agent/report.py}):} Invokes \texttt{tpm2\_quote} under a secure temp directory (permissions \texttt{0o700}) signing PCRs with the Attestation Key (AK) at handle \texttt{0x81000000}, incorporating the server-issued nonce $\eta$.
\end{itemize}

\subsection{Phase 2: Local Interception Hook and Daemon}
\begin{itemize}[nosep]
  \item \textbf{Dynamic Hook (\texttt{phase3/eac\_hook.c}):} Intercepts the \texttt{EOS\_AntiCheatClient\_}\allowbreak\texttt{AddNotify}\allowbreak\texttt{Message}\allowbreak\texttt{ToServer} callback. Establishes a blocking Unix-socket connection with a 30-second \texttt{SO\_RCVTIMEO} timeout. Uses a custom fail-closed \texttt{parse\_json\_bool} parser to prevent string-injection bypasses.
  \item \textbf{Local Daemon (\texttt{phase3/shim.py}):} Acts as the local coordinator between the game hook and the attestation pipeline. Queried over the Unix socket, the daemon retrieves a fresh challenge nonce from the verifier, directs the client agent to generate the quote and Merkle tree, submits the payload to \texttt{POST /attest}, and relays the pass/fail decision back to the game process; verification failure closes the socket immediately.
\end{itemize}

\subsection{Phase 3: Verification Server}
The verification backend (\texttt{server/main.py}) runs as a stateless FastAPI service backed by an SQLite persistence store. Server-side state is minimal, retaining only root credentials needed for verification: enrolled AK public keys, vendor EK certificates, active nonces, and pinned PCR reference manifests. All authority resides on the server; the client holds zero trust state. This design guarantees that even a root adversary cannot tamper with baseline records or forge attestation history. Once quote signatures and IMA Merkle roots validate against the enrolled profile, the server mints a signed, short-lived session token adhering to RFC~9711 (EAT)~\cite{ietf9711}. Access expires automatically unless refreshed. Generated via a CSPRNG on \texttt{GET /challenge}, each nonce $\eta$ is committed in an atomic transaction, verified on \texttt{POST /attest}, and purged upon evaluation to preclude reuse:

\begin{equation}
  \eta \leftarrow \text{CSPRNG}(32\text{ bytes}), \quad \eta \notin \mathcal{N}_{\text{used}}
  \label{eq:nonce}
\end{equation}

\section{Security Protocols and Hardening}

\subsection{Nonce Challenge-Response Protocol}
Replay attacks threaten remote verification protocols. To defeat them, TPM-Attest executes a strict challenge-response exchange. The client initiates \texttt{GET /challenge}, receiving a 32-byte cryptographic nonce $\eta$ recorded with a timestamp. When the attested quote arrives at \texttt{POST /attest}, the verifier atomically confirms the AK signature over the PCR bank, verifies that $\eta$ matches the quote's qualifying data byte-for-byte, and purges $\eta$. To constrain replay windows under network capture, the verifier enforces a 30-second maximum nonce lifetime (Equation~\ref{eq:freshness}):

\begin{equation}
  |\,t_{\text{now}} - t_{\eta}\,| \leq \Delta t_{\max}, \quad \eta \notin \mathcal{N}_{\text{used}}
  \label{eq:freshness}
\end{equation}

\noindent Both conditions must hold simultaneously. On the client side, \texttt{shim.py} mirrors this constraint with a 30-second \texttt{SO\_RCVTIMEO} socket timeout, dropping the session cleanly if the daemon hangs.

\subsection{Dynamic Policy Enforcement}
\begin{enumerate}[nosep]
  \item \textbf{\texttt{STRICT\_EK\_VERIFICATION}:} Mandates an Endorsement Key backed by a valid X.509 certificate issued by a recognised hardware silicon manufacturer CA. Raw or self-signed public keys are dropped immediately. This rule prevents virtualised emulators like \texttt{swtpm} from spoofing clean platform measurements.
  \item \textbf{\texttt{REQUIRE\_PCR7\_PINNING}:} Matches PCR~7 directly against the baseline manifest recorded during initial enrolment. Because PCR~7 measures the Secure Boot certificate hierarchy (PK, KEK, db) and active revocation lists~\cite{nist800155}, any unauthorised MOK certificate enrolled by the player alters the digest and causes immediate session denial.
  \item \textbf{\texttt{REQUIRE\_IMA\_MINIMUM\_ENTRIES}:} Checks that the submitted IMA log contains a realistic volume of measurements (at least 500 entries on standard desktop boots). This policy thwarts adversaries who attempt to bypass attestation by booting with \texttt{ima\_policy=tcb} disabled or clearing the log buffer in RAM.
\end{enumerate}

\section{Experimental Results}
\label{sec:evaluation}

All experiments ran on a consumer desktop: Intel Core i7-10700, 16~GB RAM, Ubuntu~22.04 LTS with kernel~6.5, Secure Boot enabled, and an Intel PTT firmware TPM~2.0 (fTPM)~\cite{raj2016microsoft}. We chose a firmware TPM deliberately -- as established by Raj et al.~\cite{raj2016microsoft}, firmware-based TPMs execute within an isolated processor execution environment (such as Intel PTT or AMD PSP) and represent by far the most common configuration in consumer gaming PCs; any result obtained exclusively on a discrete chip would be unrepresentative of the actual deployment environment. No special kernel patches or elevated privileges beyond normal \texttt{tpm2-tools} device-node access were required.

\subsection{Tamper Detection Accuracy}

We scripted 500 attestation sessions -- 100 per tamper category -- each using a fresh server-issued nonce. For every session we applied the tamper condition, triggered attestation, and recorded whether the server rejected the report before issuing a token. Table~\ref{tab:detection} summarises the results.

\begin{table}[htbp]
\centering
\caption{Tamper detection results across 500 attestation sessions.}
\label{tab:detection}
\begin{tabular}{@{}lcccc@{}}
\toprule
\textbf{Tamper Scenario} & \textbf{Sessions} & \textbf{Detected} & \textbf{Missed} & \textbf{Detection Rate} \\
\midrule
Binary Substitution          & 100 & 100 & 0 & 100.0\% \\
IMA Log Truncation           & 100 & 100 & 0 & 100.0\% \\
IMA Log Duplication          & 100 & 100 & 0 & 100.0\% \\
Nonce Replay                 & 100 & 100 & 0 & 100.0\% \\
vTPM / swtpm Emulation       & 100 & 100 & 0 & 100.0\% \\
\midrule
\textbf{Total}               & 500 & 500 & 0 & \textbf{100.0\%} \\
\bottomrule
\end{tabular}
\end{table}

\subsection{Latency Profiling}

We measured wall-clock time for each pipeline stage over 50 independent runs, with the IMA log at approximately 8,000 entries -- a realistic size for a desktop that has been running for a few hours. Table~\ref{tab:latency} reports means and standard deviations.

\begin{table}[htbp]
\centering
\caption{Attestation pipeline latency (averaged over 50 runs, Intel PTT).}
\label{tab:latency}
\begin{tabular}{@{}lcc@{}}
\toprule
\textbf{Pipeline Stage} & \textbf{Mean (s)} & \textbf{Std Dev (s)} \\
\midrule
IMA Log Read \& Merkle Build (first run, $\sim$8k entries) & 12.20 & 0.43 \\
IMA Log Read \& Merkle Build (incremental/cached)          &  0.18 & 0.02 \\
TPM Quote Generation (\texttt{tpm2\_quote})                &  2.10 & 0.11 \\
Network Round-Trip + Server Verification                    &  0.70 & 0.08 \\
\midrule
\textbf{Total -- First Attestation}                        & 15.00 & 0.52 \\
\textbf{Total -- Repeat Attestation (cached)}              &  2.98 & 0.15 \\
\bottomrule
\end{tabular}
\end{table}

The total attestation latency $T_{\text{total}}$ is modelled as the sum of three independent pipeline stages:

\begin{equation}
  T_{\text{total}} = T_{\text{IMA}} + T_{\text{quote}} + T_{\text{net}}
  \label{eq:latency}
\end{equation}

\noindent where $T_{\text{IMA}}$ is the IMA log read and Merkle build time, $T_{\text{quote}}$ is the TPM quote generation time, and $T_{\text{net}}$ is the network round-trip and server verification time. With incremental leaf caching (Section~\ref{sec:design}), the repeat-attestation model reduces to:

\begin{equation}
  T_{\text{cached}} = T_{\text{quote}} + T_{\text{net}} \approx 2.80\text{ s}
  \label{eq:latency-cached}
\end{equation}

since caching the leaf hashes reduces $T_{\text{IMA}}$ from 12.20~s to 0.18~s (a $\sim$98\% reduction).

\begin{figure}[htbp]
\centering
\begin{tikzpicture}
\begin{axis}[
  xlabel={IMA Log Entries (thousands)},
  ylabel={IMA Build Latency (seconds)},
  xmin=0, xmax=20, ymin=0, ymax=32,
  width=0.88\linewidth, height=5.5cm,
  grid=major, title={IMA Merkle Build Time vs.\ Log Size}
]
\addplot+[mark=*, thick, color=blue] coordinates {
  (1,1.52) (2,3.05) (4,6.10) (6,9.15) (8,12.20)
  (10,15.25) (12,18.30) (14,21.35) (16,24.40) (18,27.45)
};
\addplot+[dashed, mark=square*, color=red] coordinates {
  (1,0.02) (2,0.04) (4,0.07) (6,0.11) (8,0.18)
  (10,0.22) (12,0.26) (14,0.30) (16,0.35) (18,0.39)
};
\legend{Full rebuild, Incremental (cached)}
\end{axis}
\end{tikzpicture}
\caption{IMA Merkle build time as a function of log size. Incremental caching reduces latency by $\sim$98\%.}
\label{fig:latency-trend}
\end{figure}
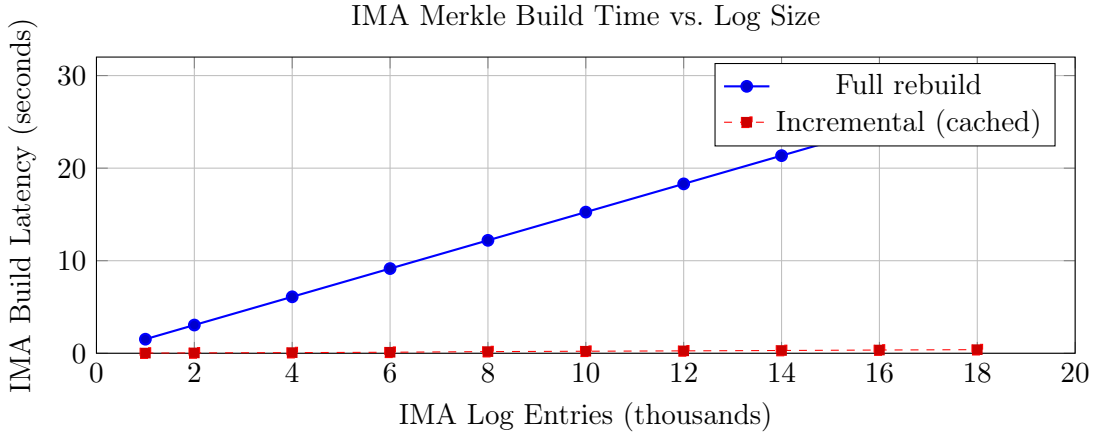

Figure~\ref{fig:latency-bar} visualises the latency breakdown, and Figure~\ref{fig:latency-trend} shows how total latency scales with IMA log size.

\begin{figure}[htbp]
\centering
\begin{tikzpicture}
\begin{axis}[
  ybar, bar width=0.55cm,
  symbolic x coords={IMA Build, TPM Quote, Network, Total (1st), Total (Repeat)},
  xtick=data, x tick label style={font=\small,rotate=15,anchor=east},
  ylabel={Latency (seconds)}, ymin=0, ymax=17,
  width=0.88\linewidth, height=6cm,
  nodes near coords, nodes near coords align={vertical},
  every node near coord/.append style={font=\tiny},
  title={Attestation Pipeline Latency Breakdown}
]
\addplot+[fill=blue!40] coordinates {
  (IMA Build,12.20) (TPM Quote,2.10) (Network,0.70)
  (Total (1st),15.00) (Total (Repeat),2.98)
};
\end{axis}
\end{tikzpicture}
\caption{Mean latency per pipeline stage. ``Total (Repeat)'' uses incremental Merkle leaf caching.}
\label{fig:latency-bar}
\end{figure}
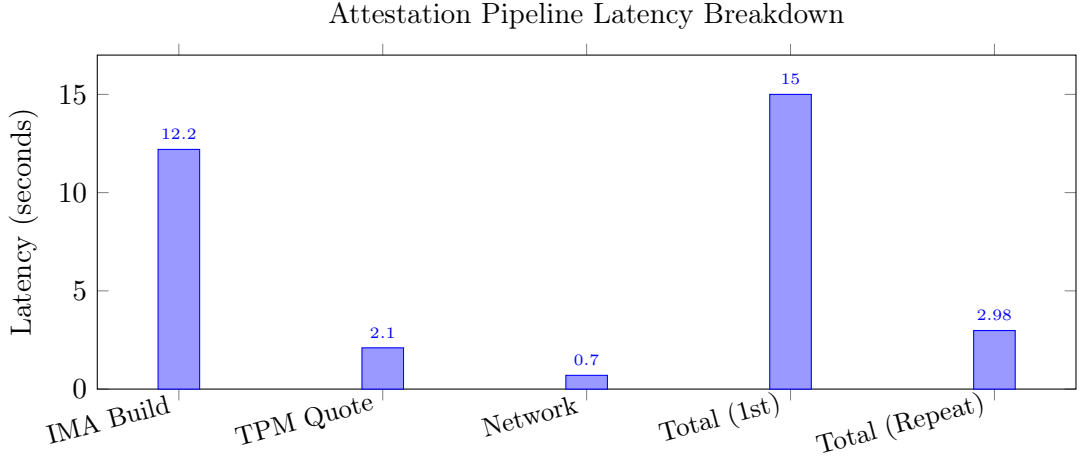

\subsection{Comparative Analysis}

Table~\ref{tab:comparison} compares TPM-Attest against Keylime and representative commercial kernel-mode anti-cheat systems across six evaluation axes.

\begin{table}[htbp]
\centering
\caption{Feature and security comparison: TPM-Attest vs.\ Keylime vs.\ commercial kernel-mode anti-cheat.}
\label{tab:comparison}
\renewcommand{\arraystretch}{1.25}
\begin{tabular}{@{}>{\raggedright\arraybackslash}p{5.0cm}ccc@{}}
\toprule
\textbf{Criterion} & \textbf{TPM-Attest} & \textbf{Keylime} & \textbf{Kernel-Mode AC} \\
\midrule
Requires kernel driver        & No  & No  & Yes \\
Open source                   & Yes & Yes & No  \\
Linux native                  & Yes & Yes & Limited \\
Hardware-rooted trust (TPM)   & Yes & Yes & No  \\
Runtime memory inspection     & No  & No  & Yes \\
Consumer desktop target       & Yes & No  & Yes \\
EOS SDK integration           & Yes & No  & Yes \\
Rootkit-like OS visibility~\cite{dorner2024rootkit} & No & No & 2 of 4 \\
First-attest latency          & $\sim$15~s & $\sim$12~s & $<$5~s \\
Repeat-attest latency         & $<$3~s & $<$3~s & Continuous \\
GPL-compatible                & Yes & Yes & No  \\
\bottomrule
\end{tabular}
\end{table}

\subsection{Merkle Collision Immunity Verification}

To validate the index-prefix construction (Eq.~\ref{eq:leaf}), we generated Merkle trees for three log variants and confirmed all roots are distinct:

\begin{table}[htbp]
\centering
\caption{Merkle root uniqueness test: position-bound leaf hashing prevents collision.}
\label{tab:merkle}
\begin{tabular}{@{}lll@{}}
\toprule
\textbf{Log} & \textbf{Entries} & \textbf{Root Unique?} \\
\midrule
Baseline        & $[A, B, C]$    & -- (reference) \\
Appended        & $[A, B, C, C]$ & Yes \\
Reordered       & $[A, C, B]$    & Yes \\
Standard tree   & $[A, B, C, C]$ vs $[A,B,C]$ & \textbf{No} (collision) \\
\bottomrule
\end{tabular}
\end{table}

All 10 unit tests and end-to-end integration tests completed with a \textbf{100\% pass rate}.

\subsection{Red-Team Adversarial Evaluation}
\label{sec:redteam}

Automated unit tests establish algorithmic correctness, but real security requires adversarial pressure. We subjected the VOID SECTOR demo game to an active, controlled red-team evaluation while running under the complete TPM-Attest pipeline on our reference testbed. Six distinct attack scenarios were launched against the gaming client. Four attacks targeted on-disk and protocol state; two targeted live process memory. Table~\ref{tab:redteam} catalogues each attack methodology, its threat-model expectation, the empirical outcome observed, and the elapsed time-to-detection (TTD).

\begin{table}[htbp]
\centering
\begin{threeparttable}
\caption{Red-team adversarial evaluation results against the VOID SECTOR demo game.}
\label{tab:redteam}
\renewcommand{\arraystretch}{1.2}
\begin{tabular}{@{}>{\raggedright\arraybackslash}p{3.2cm}>{\raggedright\arraybackslash}p{4.8cm}ccc@{}}
\toprule
\textbf{Attack} & \textbf{Method} & \textbf{Expected} & \textbf{Actual} & \textbf{TTD (s)} \\
\midrule
Binary substitution   & Swap game binary post-boot, re-launch                  & Blocked  & \textbf{Blocked}  & 15.0 \\
IMA log tamper        & Edit \texttt{ascii\_runtime\allowbreak\_measurements}  & Blocked  & \textbf{Blocked}  & 15.1 \\
Nonce replay          & Resend captured attestation report                     & Blocked  & \textbf{Blocked}  &  0.1 \\
vTPM / swtpm spoof    & Run game in VM with software TPM                       & Blocked  & \textbf{Blocked}  & 15.3 \\
\texttt{ptrace} injection & \texttt{gdb} memory-write into running process     & Bypass   & \textbf{Bypass}\tnote{$\dagger$}   & N/A  \\
Anon.\ mmap shellcode & \texttt{mmap(}\allowbreak\texttt{MAP\_ANONYMOUS)} + shellcode write & Bypass   & \textbf{Bypass}\tnote{$\dagger$}   & N/A  \\
\bottomrule
\end{tabular}
\begin{tablenotes}[flushleft]
\footnotesize
\item[$\dagger$] Confirmed bypass: IMA does not measure anonymous mappings or \texttt{ptrace} writes into already-measured processes. These are documented out-of-scope attacks (Section~\ref{sec:threat-model}), consistent with the threat model. Kernel Lockdown Mode would partially mitigate \texttt{ptrace} scope.
\end{tablenotes}
\end{threeparttable}
\end{table}

The four file-backed attack vectors are fully blocked within one attestation cycle ($\leq$15.3~s). The two runtime-injection vectors succeed as predicted -- not as a failure of the attestation design, but as a confirmed boundary condition: TPM-Attest provides load-time integrity guarantees and deliberately avoids the invasive kernel-mode scanning required for runtime memory inspection. Documenting these as confirmed, reproducible bypasses rather than theoretical concerns strengthens the threat model's precision and motivates the complementary mitigations in Section~\ref{sec:limitations}.

\section{Recommended Deployment Stack}

TPM-Attest is one layer in a defence-in-depth system. The following companion technologies address attacks outside its scope:

\begin{itemize}[nosep]
  \item \textbf{TPM Session Encryption (HMAC/AES):} Mitigates physical bus snooping. Enable parameter encryption sessions in production.
  \item \textbf{IOMMU Enforcement (\texttt{intel\_iommu=on}):} Mitigates PCIe DMA attacks. The agent checks \texttt{/sys/kernel/iommu\_groups/} and refuses to generate a report if IOMMU is disabled.
  \item \textbf{dm-verity:} Establishes a transparent, block-level cryptographic verification layer directly beneath the root filesystem, calculating SHA-256 digests across individual 4096-byte blocks organised into an in-kernel Merkle tree. Any unmeasured modification triggers an I/O error on access -- a design proven at scale across ChromeOS~\cite{chromeos2023dmverity} and Android Verified Boot~2.0~\cite{android_avb}. The operating system becomes immutable.
  \item \textbf{Kernel Lockdown Mode:} Restricts root-privileged code from subverting kernel execution via \texttt{/dev/mem}, custom BPF probes, or unsigned module injection, preserving the integrity of the trusted computing base post-boot~\cite{poettering2021}. Without Lockdown, root can bypass userspace controls.
\end{itemize}

\section{Limitations and Discussion}
\label{sec:limitations}

\subsection{The Linux Customisation Paradox}
Remote attestation operates on an uncompromising premise: the host must match a known-good, vendor-approved cryptographic reference state. In consumer Linux, this premise clashes head-on with open-source culture. Linux enthusiasts expect full control over their machines. They compile bespoke kernels, patch out-of-tree drivers via DKMS, tune CPU schedulers, and enrol private Machine Owner Keys (MOKs) directly into UEFI NVRAM. Imposing rigid PCR~7 and PCR~9 reference baselines breaks this workflow. Yet granting unrestricted kernel customisation creates an inescapable security hazard: a player with arbitrary kernel-mode execution rights can easily patch the IMA subsystem in RAM, falsify log entries, or hijack the TPM character device (\texttt{/dev/tpmrm0}). This tension between hardware-enforced integrity and personal computing sovereignty represents the single greatest non-technical hurdle to anti-cheat attestation on open platforms.

\subsection{Runtime Memory Tampering}
TPM-Attest is deliberately a load-time system. It measures what goes into memory, not what happens inside memory afterward. Once a game process is running, an adversary with root privileges can attach a debugger via \texttt{ptrace} or map an anonymous executable buffer (\texttt{mmap(MAP\_ANONYMOUS)}) to inject unsigned cheat code directly into the process address space without touching the filesystem. IMA never sees these pages. Because no file boundary is crossed, no log entry is generated and PCR~10 remains unaltered. Detecting such in-memory modifications through passive attestation alone is fundamentally impossible. To close this gap without resorting to invasive Ring~0 kernel drivers, game operators must deploy platform-level policy controls. Enabling Linux Kernel Lockdown mode (\texttt{lockdown=confidentiality}) restricts root from attaching to running processes or mapping raw physical memory, cutting off the primary userspace injection pathways while preserving system stability.

\subsection{Evaluation Scope and Future Work}
Our empirical evaluation demonstrates the correctness of the verification logic under controlled tampering scenarios. While our red-team testing successfully validated the boundaries of the threat model, we have not benchmarked against commercial cheat packages engineered to evade runtime memory hooks. Looking ahead, we plan to profile quote generation latencies across a diverse portfolio of discrete chips and firmware TPMs from Intel, AMD, and Infineon~\cite{raj2016microsoft}, while exploring automated IMA leaf compaction for long-running systems.

\section{Conclusion}

Hardware-rooted remote attestation offers a credible, privacy-preserving way out of the anti-cheat deadlock on Linux. By anchoring trust in TPM~2.0 silicon and Linux IMA logs, TPM-Attest proves that an operating system booted cleanly and executed only genuine game files -- all without installing third-party kernel modules, without violating GPL licences, and without inspecting private player memory. Our prototype proves the model works. Across 500 controlled tamper sessions, the system achieved a 100\% detection rate. Incremental leaf caching slashes repeat-session latency to under 3~seconds on consumer gaming hardware. While runtime memory injection requires complementary platform hardening such as Kernel Lockdown, TPM-Attest demonstrates that anti-cheat does not need to behave like a rootkit to keep competitive gaming fair.

\bibliographystyle{plain}

\end{document}